\pdfoutput=1
\documentclass[twocolumn]{aastex61}

\received{hopefully soon}
\revised{hopefully soon after}
\accepted{2017-10-05}

\begin{document}

\title{MARXS: A modular software to ray-trace X-ray instrumention}

\correspondingauthor{Hans Moritz G\"unther}
\email{hgunther@mit.edu}

\author[0000-0003-4243-2840]{Hans Moritz G\"unther}
\affiliation{MIT, Kavli Institute for Astrophysics and Space Research,
77 Massachusetts Avenue, Cambridge, MA 02139, USA}

\author{Jason Frost}
\affiliation{Department of Physics, Stanford University, Stanford, California 94305, USA}
\affiliation{MIT, Kavli Institute for Astrophysics and Space Research,
77 Massachusetts Avenue, Cambridge, MA 02139, USA}
\author{Adam Theriault-Shay}
\affiliation{MIT, Kavli Institute for Astrophysics and Space Research,
77 Massachusetts Avenue, Cambridge, MA 02139, USA}

\begin{abstract}

 To obtain the best possible scientific result, astronomers must understand the
 properties of the available instrumentation well.  This is important both when
 designing new instruments and when using existing instruments close to the
 limits of their specified capabilities or beyond.  Ray-tracing is a technique
 for numerical simulations where the path of many light rays is followed through
 the system to understand how individual system components influence the
 observed properties, such as the shape of the point-spread-function (PSF).  In
 instrument design, such simulations can be used to optimize the performance. 
 For observations with existing instruments this helps to discern instrumental
 artefacts from a true signal.  Here, we describe MARXS, a new python package
 designed to simulate X-ray instruments on satellites and sounding rockets.
 MARXS uses probability tracking of photons and has polarimetric capabilities.

\end{abstract}


\section{Introduction}
\label{sec:introduction}
Astronomical observations rely on increasingly complex instruments always
pushing the boundaries in spatial and spectral resolution. Simulation software
is an important tool to help astronomers and engineers to design and operate
these instruments, to propose and plan observations, and to analyze data to
extract scientific results.

Depending on the application, simulations with different levels of detail are
needed. For example, to estimate if two close sources seen by one instrument
will also be resolved by another, it might be sufficient to convolve an image
with the point-spread-function (PSF) of the second instrument. On the other
hand, for instruments where the PSF is strongly dependent on the wavelength,
where the PSF changes significantly over the field-of-view (FOV), or where the
pointing direction changes within an observation (e.g. due to dithering), more
complex simulation tools are required. One approach is to perform a geometrical
ray-trace simulation, where rays are propagated through all components of the
instrument. The simulation starts at some position $\vec r_0$ (typically chosen
in the aperture plane) with a ray direction $\vec d_0$ (parallel to the optical
axis of the telescope for on-axis sources). The simulation follows the ray and
calculates where it next intersects an element of the instrument, e.g. it might
hit a mirror surface at position $\vec r_1 = \vec r_0 + \lambda \vec d_0$. At
this point, the new propagation direction $\vec d_1$ of the ray is chosen. For
a mirror surface, this depends on the angle of incidence, but other optical
elements such as diffraction gratings or polarizing filters might take other
ray properties into account. In this way the simulation tracks the ray until it
hits (or misses) the detector. The process is repeated for many rays starting
at different $r_0$ and with different wavelengths and polarization vectors. In the
end, the list of rays can be analyzed. For example, the position of
intersection with the detector can be binned into a 2-D histogram to yield a
simulated detector image. Ray-tracing as a technique is particularly
well-suited to track the influence of aberrations such as coma and astigmatism,
to predict the shadows cast by support structures and baffles, and to analyze
the effect of gratings and detectors that are flat, while the analytically
derived focal plane would require them to be curved.

Ray-tracing helps with the design of instruments
\citep[e.g.][]{2014SPIE.9144E..2EW} when decisions about the placement of the
instrument components are made, it helps to predict the performance of
instruments, can be used to derive calibration products such as the
instrumental line width of a spectrograph \citep{2000SPIE.4140..559F}, and
finally it can be used predict the instrumental signal for a specific
astrophysical model to facilitate proposal writing or for comparison to
observed data \citep[e.g.][]{2009ApJ...699.1004Z}.

To give a few examples how ray-tracing helps to enable specific scientific
results, the MARX software
\citep{2012SPIE.8443E..1AD}\footnote{\url{http://adsabs.harvard.edu/abs/2013ascl.soft02001W}}
can generate a model of the PSF of the \emph{Chandra} observatory at a
specific position in the focal plane for the specific
dither pattern and source spectrum of the observation in question. Using this,
\citet{2012MNRAS.422..590R} are able to model the wings of the PSF in the
quasar PKS 1229-021 and detect an extended emission component from the
cool-core cluster surrounding it and similarly \citet{2017MNRAS.468..109W} find
extended emission around the radio galaxy C~41.17. In a similar use the
ACISExtract package \citep{2010ApJ...714.1582B}\footnote{\url{http://adsabs.harvard.edu/abs/2012ascl.soft03001B}} runs a MARX
simulation for every detected point source to determine the appropriate
extraction radius. The resulting flux scale is directly tied to the accuracy of
the ray-trace simulation. ACISExtract is often used in the analysis of stellar
clusters \citep[e.g.][]{2013ApJS..209...27K}. \citet{2014ApJ...788...53M} used
MARX simulations to estimate the uncertainty in the zero-point of the velocity
scale in a dispersion grating to verify the redshift of absorption lines
observed in the transient X-ray binary MAXI J1305-704.

In this paper, we present a new python software package called MARXS
(Multi-Architecture-Raytrace-Xraymission-Simulator) that derives many ideas
from the \emph{Chandra} MARX code. Previous ray-trace codes are often either
specific to a single X-ray mission like MARX for \emph{Chandra} or SciSim
\citep{2005SPIE.5898..450G} for
\emph{XMM-Newton} or they are not easily available
to all potential users of a mission because they rely on expensive commercial
packages or are implemented in very specialized programming languages like
MT\_RAYOR \citep{2011SPIE.8147E..0YW} which is implemented in Yorick and used
for {NuStar} and \emph{ASTROSAT}. Our new MARXS package is written in Python
and distributed under version 3 of the GNU Public license (GPL). MARXS provides
generic implementations of elements commonly found in X-ray instruments which
can be combined as needed to setup simulations for a wide range of use cases.

In section~\ref{sect:design} we explain the design principles and capabilities
of the MARXS code. In section~\ref{sect:example} we show examples of using
it.
We end with a short summary in section~\ref{sect:summary}. This paper describes
version 1.1 of MARXS \citep{marxs1.1}.
Development continues on github\footnote{\url{https://github.com/chandra-marx/marxs}}.

\emph{Note to astroph readers: This article was accepted by AJ and the journal will host interactive versions of figures 1 and 3 after publication. In the meantime, interactive versions of those figures can also be found in the documentation of the MARXS software. The specific url is given in the figure caption.}

\section{Design and capabilities}
\label{sect:design}
This section explains aspects of the software design and mentions some of
the modules and classes in MARXS. Module names, classes and other Python code
are written \texttt{monospaced}. This section is not intended
to replace the MARXS documentation\footnote{\url{http://marxs.readthedocs.io/}}
where further details and code examples can be found.

MARXS contains classes for X-ray sources (in module \texttt{marxs.source}) and
optical elements such as mirrors, baffles, diffraction gratings, and detectors
(\texttt{marxs.optics}). The \texttt{marxs.simulator} modules contains classes
that help grouping optical elements, e.g.\ different CCDs that make up one
camera. There are supporting modules to help with the instrument design in
\texttt{marxs.design} to calculate the proper placement of diffraction
gratings and detectors on a torus for a Rowland spectrometer \citep[see
  e.g.][]{Beuermann:78}. The display of results is handled in
(\texttt{marxs.visualization}), as well as utility modules for mathematics.

\subsection{Order of elements in the simulation}
MARXS visits elements sequentially. After a ray passes (and potentially
interacts with) element $i$, MARXS tests if the ray interacts with element
$i+1$. The ray never returns to element $i$. This approach reduces the number
of intersection tests that need to be calculated per ray-trace, but it requires
an A-priory knowledge about the order of elements that a ray might intersect
with. This means that MARXS is of limited use to estimate the scattered light in an instrument, because it cannot handle rays that bounce back to ``earlier'' elements.

\subsection{Units and coordinate systems}
MARXS uses homogeneous coordinates, which describe position and direction in a
4 dimensional coordinate space [$x$, $y$, $z$, $w$]. For example, [3,0,0,1] describes a point at
$x/w=3$, $y/w=0$, and $z/w=0$. The vector [3,0,0,0] describes a point at infinity, because $x/w$=3/0. Points at infinity can be thought of a ``direction vector'', so that a [1,0,0,0], [2,0,0,0] or [3,0,0,0] would all describe a vector parallel to the x-axis.
In homogeneous coordinates, rotation, zoom, and translations together can
be described by a [4,4] matrix and several of these operations can be chained
simply by multiplying the matrices.

Every optical element in MARXS has such a [4,4] matrix associated with it in an
attribute called \texttt{pos4d}. We experimented with astropy's unit system
\citep{2013A&A...558A..33A}, but found that it added significant
overhead. Thus, dimensional numbers are represented as simple floating point
numbers in the code, and by convention we assume all spatial units to be given
in mm and all energies in keV. Use of astropy's units system for input with
automatic conversion to these base units internally is planned for a future
version.

\subsection{Rays and photons}
The fundamental unit of ray-tracing is a single ``ray'', which can be
interpreted as a wave package of many photons or as a single
photon. Most X-ray detectors are photon counting, thus it is
convenient to identify each ray with a single photon in this case.
MARXS generates a list of photons, which is stored in an
\texttt{astropy.table.Table} object \citep{2013A&A...558A..33A}. This object
holds numpy arrays and associated metadata. It also provides the capability
of saving data in fits and other formats, while at the same time, mathematical
operations such as dot and cross-products can be performed efficiently on the
numpy arrays.

\subsection{No simulation of microphysics}
Currently, MARXS does not provide classes to calculate the result of an
interaction with an element from the properties of the material. Instead, it
requires the user to provide look-up tables for quantities such as reflection
probabilities or grating efficiencies. Alternatively, the user can supply code
to calculate these quantities from first principles.

As an example, the MARXS class \texttt{FlatGrating} does not
calculate the diffraction probability in every order from the optical constants
of the materials that make up the gratings and the dimensions of individual
bars; instead, it uses tabulated diffraction efficiencies from a file and
assigns a diffraction order to every ray based on those numbers, so that the
dispersion angle can be calculated from the grating constant and the grating
equation. This approach allows the flexibility to use either theoretical
grating efficiencies pre-calculated by some other program or interpolate
experimentally determined values.

\subsection{Polarization}
MARXS supports polarization ray-tracing, but the proper treatment of
polarization is not yet implemented for all components of MARXS.

The polarization of each ray is represented as a 3d-vector and uses matrices to
move this vector in space. As an extension of the 2-d Jones calculus, this
method is particularly suited for light paths that are not all parallel
\citep{1992SPIE.1746...62C}. See \citet{Yun:11} and \citet{Yunthesis} for
details on polarization ray tracing and the derivation of the relevant
matrices.

This mechanism can handle both linear and circular polarized light. However,
the light sources currently included in MARXS only support linear polarization.

\subsection{Sources and apertures}
The module \texttt{marxs.sources} provides several classes for point sources
both in the laboratory and on the sky. There are also a few classes for
spatially extended celestial sources such as circle, disk, or Gaussian
luminosity distribution. All sources allow great freedom to specify spectral
properties, polarizations, and timing behavior through keywords. For
astrophysical sources, the telescope pointing must be specified to transform
the origin of the photons in sky coordinates to a direction in the instrument
coordinate system. The aperture class then assigns the ray a randomized initial
position in the instrument frame. This random sampling of the aperture means
that simulations need to be run with a number of rays that is sufficient to
sample the aperture.

Laboratory sources generate rays in the coordinate system of the experiment and
no further pointing or aperture is required.

\subsection{Parts and pieces}
MARXS provides a range of generic optical elements such as mirrors, gratings,
filters, and detectors in \texttt{marxs.optics}. Each object is initialized
with a 4-d matrix that sets positions and orientation compared relative to the global coordinate system, and the size of
each element. Additional properties such as pixel size or grating constant are
passed in as keywords. Some of these elements are demonstrated in the examples
below.

The optical elements are laid out in a hierarchy based on inheritance. This makes
is easy to add new elements, since they can inherit most of their functionality
from existing base classes, so that only little new code has to be written.

\subsection{Probability tracking}
Many elements in an X-ray optics absorb photons. One way to treat this in a
Monte-Carlo ray-trace is to draw a random number and decide for each ray if it
should be deleted from the list of rays or propagated further. However, in the
setup of MARXS it is much easier to operate on a photon list of fixed
size. Thus, each photon in MARXS has a \texttt{probability} associated with
it. At the source, the value $1.0$ is assigned and in any absorbing element
this number is decreased to track the probability that the photon is still
present. For example, a photon reflected by a mirror with 80\% reflection
efficiency and passing through an optical blocking filter with 90\% X-ray
transmission, will have $p=0.8*0.9 = 0.72$.

\subsection{Visualization}
MARXS photons lists are just astropy tables, and as such they have built-in
capability to be written to a variety of file formats, including fits
tables. Thus, any standard astronomy software can be used to analyze simulation
results. For example, binning the x and y coordinates of the rays on the
detector yields a detector image. Furthermore, MARXS also provides several ways
to pass the coordinates of the instrument components and of the rays to 3-d
visualization tools. Currently supported are Mayavi \citep{mayavi}, a Python
library, and three.js, a javascript library. Both of them make use of the
OpenGL standard, so that the 3-dimensional rendering is done in the graphics
card of the computer, making fluid animations of instrument designs with
hundreds of mirror modules, gratings and detectors and thousands of photon
pathways possible. In particular, Mayavi can export the data to the x3d format,
which is particularly suited for inclusion in presentations and publications
\citep{2016ApJ...818..115V}. Figures~\ref{fig:subaper3d} and \ref{fig:3dpol}
are produced in this way and are fully interactive in the online version of
this article.

\subsection{Verification}
MARXS employs continuous integration with a few hundred unit tests to reduce
the risk of unintended regressions. Test cases include
verification that all classes adhere to the common interface (e.g.\ all optical
elements can be called with a photon list as argument), physical boundary
conditions are preserved (e.g.\ the polarization vectors are always
perpendicular to the directions of the ray), and, where available, analytical
test cases are correctly simulated (e.g.\ the grating equation for a
diffraction grating).

\section{Examples}
\label{sect:example}
In this section, we show two examples of experiments that can be easily simulated  with MARXS. Both examples are covered in more detail in the documentation of MARXS\footnote{\url{http://marxs.readthedocs.io/}} where the full source code is given.

\subsection{The benefit of sub-aperturing}
Collimation of X-rays typically requires double reflection of a mirror
shell at grazing incidence. When they leave the mirror, photons are distributed
on the surface of a cone where the focal point of the mirror is the tip of the
cone. In some situations, it can be beneficial to use only a fraction of the mirror. This is called sub-aperturing. In this
example, we simulate an instrument with sub-aperturing. Instrument parameters
are inspired by the \emph{Chandra}/HETG \citep{2005PASP..117.1144C}, but
optical elements are much bigger, so that we can explore what effect large
diffraction gratings and detectors have. On the one hand, using larger elements
reduces the area covered by support structures and thus increases the
throughput of the telescope; on the other hand, large, flat elements deviate
more from the analytic curved surface where gratings and detectors should be
placed in a Rowland geometry.

\begin{figure}[ht!]
\plotone{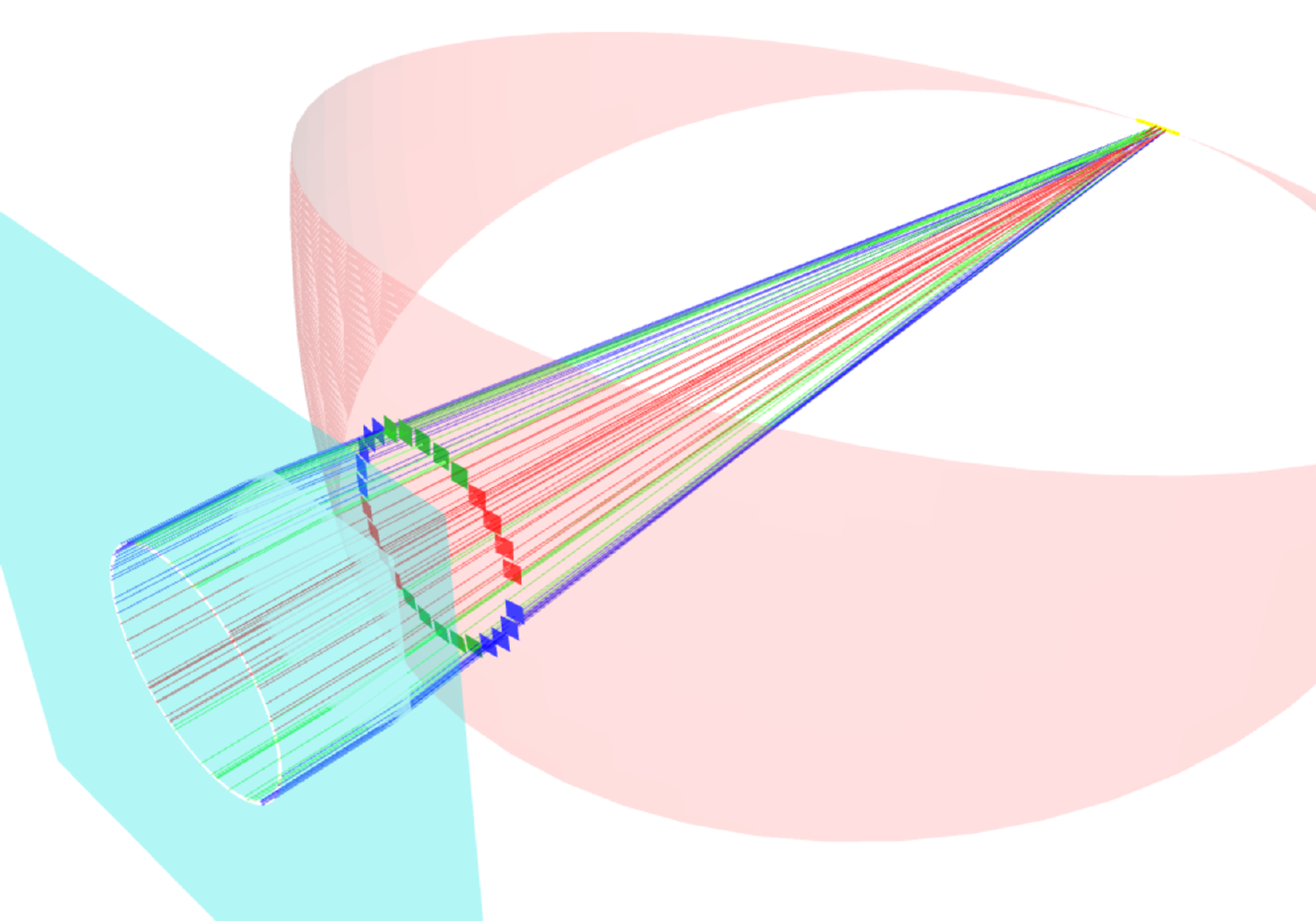}
\caption{A simulation to study sub-aperturing. The entrance aperture is a
  narrow ring in the blue plane. The gray plane indicates the simplified
  mirror model. Diffraction gratings (red, green, and blue) as well as
  detectors (yellow, in the top right corner) are placed in the Rowland Torus
  (red surface). Lines are 1~keV rays, which are colored to match the grating
  they pass through. For visibility only a few hundred rays are shown.
  \emph{An interactive version of this figure is available at \url{https://marxs.readthedocs.io/en/latest/visualization.html} (about 2/3 down the page)}.
}
    \label{fig:subaper3d}
\end{figure}

In this simulation, the focal length is 11~m, the grating constant is 200~nm,
the gratings are 160 mm on each side, and detectors are rectangular with
$100\times40$~mm$^2$. Figure~\ref{fig:subaper3d} shows a ray-trace for a source
on axis with a monochromatic flux of 1~keV. The entrance aperture is a narrow
ring in the blue, partially transparent plate. Rays are
placed in this aperture and propagated parallel to the optical axis. The mirror
is approximated with a simple model: All photons passing the gray plate in the figure are
focused to the focal point, then some random scatter is added. Figure errors,
surface roughness, and particulates scatter X-rays by a larger angle in the
plane of incidence than perpendicular to it \citep{Cash:87}. Thus, the scatter
is typically larger in the plane of incidence than perpendicular to it
\citep{1993SPIE.1742..171O,2015SPIE.9603E..0KC}. In this simulation, we
explore the case where in-plane scatter is a few times larger than scatter
perpendicular to the plane of reflection.  Transmission gratings (red, green,
and blue) and detectors (yellow) are placed on the Rowland Torus, a section of
which is shown as a partially transparent, red surface. In the figure, the
detectors are very hard to see, because they are only a few cm large and are
located several meters from the front of the telescope. This difference in
scale is often found in X-ray telescopes and thus it is very useful to be able
to interactively pan and zoom in the figure - this is possible in the
electronic version of this article.

\begin{figure}[ht!]
\plotone{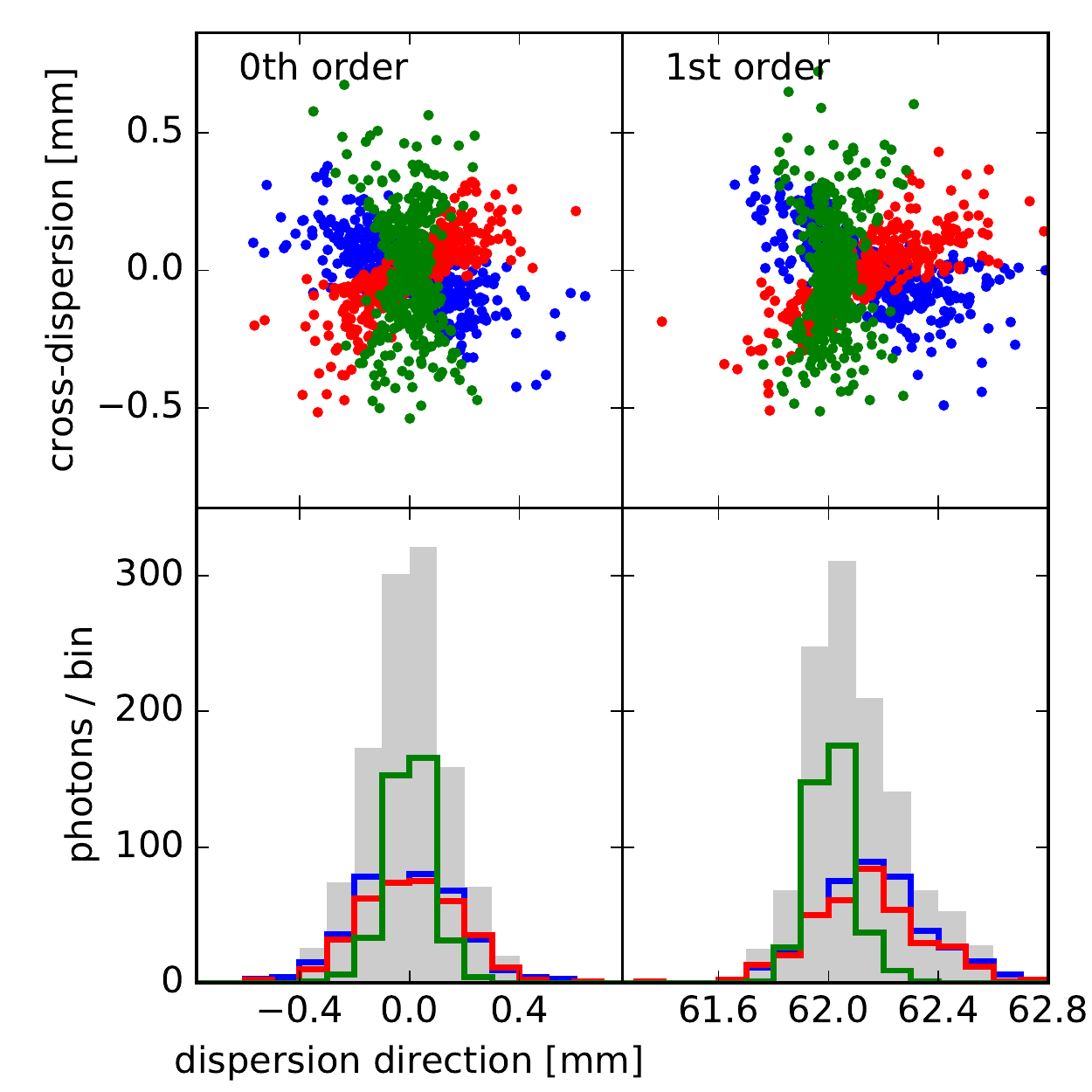}
\caption{\emph{top:} Photon positions on the detector \emph{bottom:} Histogram
  of the distribution in dispersion direction. Colors match the color of the
  dots in the top row and respective diffraction gratings in
  figure~\ref{fig:subaper3d}. The gray blocks show the summed
  histogram. \emph{left:} zeroth order. \emph{right:} first order. Note that
  this figure displays a simulation with a larger number of photons than figure~\ref{fig:subaper3d}.}
    \label{fig:subaper}
\end{figure}

Rays are colored according to the grating they pass
through and they are diffracted with equal probability into the grating orders
-1, 0, and +1. Figure~\ref{fig:subaper} also shows the
detector image of the zeroth and first order. 

The PSF in zeroth order is rotationally symmetric, but if photons are
selected through a sub-aperture, the PSF narrows considerably in one
direction. In the plot, the photons shown in green are narrow in the dispersion
direction and thus they give a sharper feature in the first order than the
other photons, leading to a spectrograph with increased resolving power. By
varying several parameters of the simulation such as the size of gratings and
detectors, MARXS can also be used to understand the asymmetric shape of the
first-order PSF, which is due to a combination of
aberration and the finite-size of the large gratings, but a detailed discussion
is beyond the scope of this simple usage example.

\subsection{The effect of polarization}
\begin{figure}[ht!]
\plotone{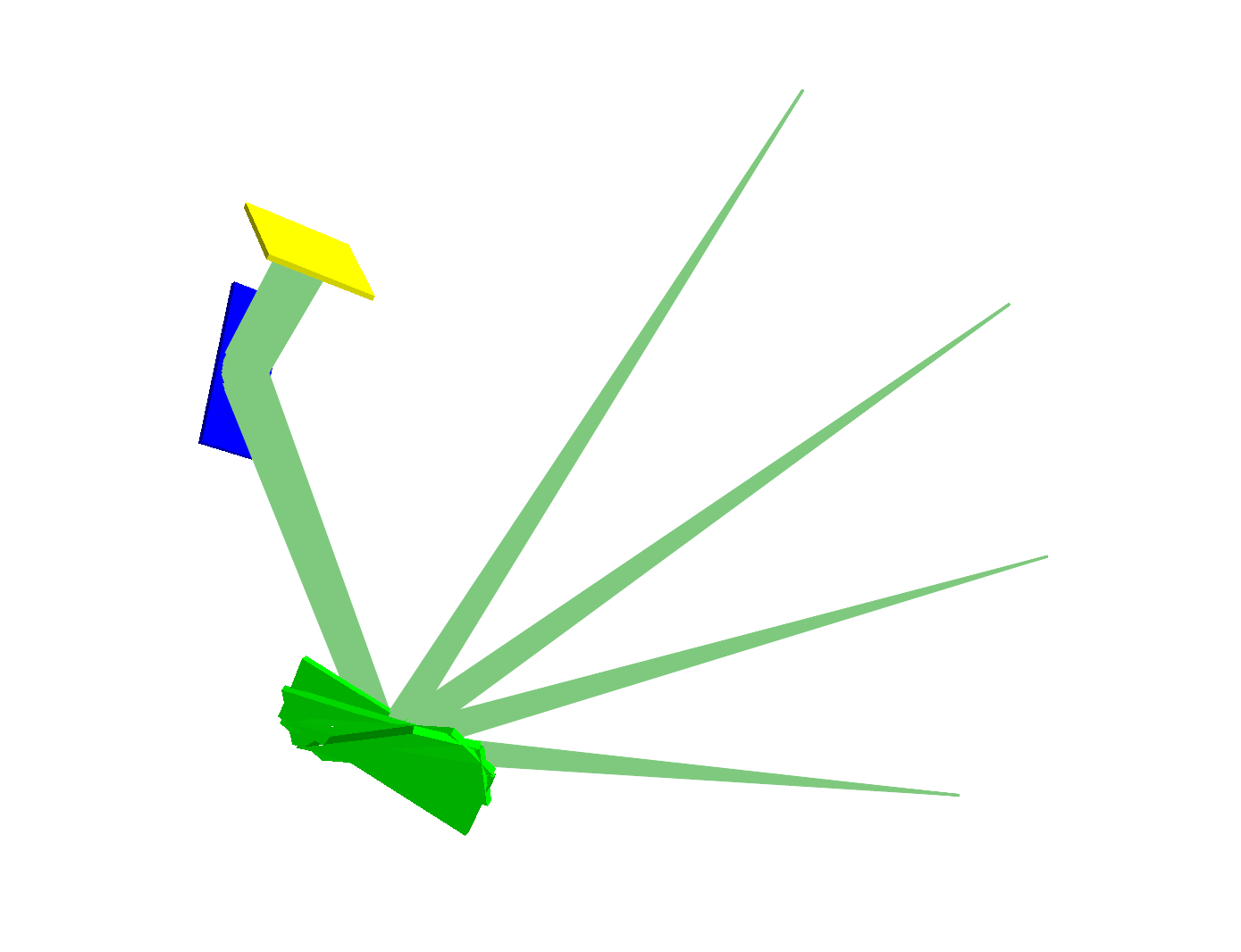}
\caption{Setup of the polarization experiment. The components are two
  mirrors (green and blue) and the detector (yellow). Simulated rays
  are shown in green, the light source is at the tip of the cone.
  The figure shows four
  positions for the lightsource and the corresponding rotation of the
  green mirror. The green mirror is positioned in such a way that all
  rays reach the blue mirror and the detector for every rotation angle
  of the source.
  \emph{An interactive version of this figure is available at \url{https://marxs.readthedocs.io/en/latest/examples.html} (about 2/3 down the page).}
  }
    \label{fig:3dpol}
\end{figure}

\begin{figure}[ht!]
\plotone{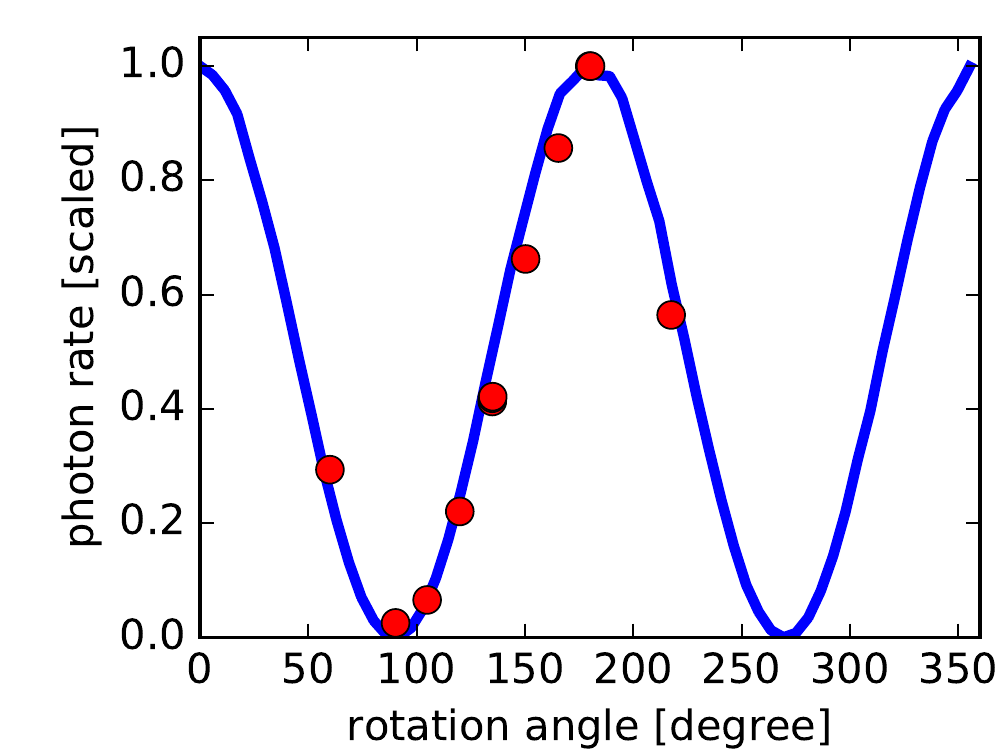}
\caption{Photon fluxes for a simulation for different rotation angles of
source and green mirror. The data points are taken from \citet{2013SPIE.8861E..1DM} where this setup is tested in the lab (physical dimensions for the mirrors and the distance between the elements differ from the simulation). Statistical} error bars are smaller than the plot symbols.
    \label{fig:polcurve}
\end{figure}

This example reproduces a setup in a laboratory X-ray beamline, where
X-ray polarization using multi-layer mirrors is studied
\citep{2013SPIE.8861E..1DM}. The experiment consists
of four elements (figure~\ref{fig:3dpol}): First, there is an X-ray source that can be operated
with a range of anodes to control the spectrum of the source. To
simplify matters for the simulation, we will assume monochromatic
light here. The source has a small opening aperture and shines on a
multi-layer mirror (green) with an incidence angle of 45$^\circ$, which matches the 
Brewster angle. This angle is defined as the angle where p-polarized light is perfectly transmitted and only s-polarized light (polarization direction
perpendicular to the plane of reflection) will be
reflected. When unpolarized light arrives at the mirror at the Brewster angle, the reflected light is thus 100\% polarized. X-rays transmitted through the reflecting surface will be absorbed by the lower layers of the mirror substrate and the holder and are lost, we only track the reflected light.  Multilayer coatings can reflect a few percent of the
incoming X-rays. Together, the source and this first mirror can be
thought of as a ``source of polarized X-rays''. From the first mirror,
X-rays are directed onto a second mirror (blue), which they reach again at the
Brewster angle, and finally a detector (yellow). Source and first mirror can be
rotated with respect to the second mirror, changing the
polarization angle of the light that reaches the second mirror. If the
incoming light is s-polarized with respect to the second mirror, it
will reflect a fraction of the light to the detector. When source and
first mirror are rotated by 90$^\circ$ the polarization vector of the
photons will be parallel to the plane of reflection with the second
mirror, and no light is reflected. Figure~\ref{fig:3dpol} shows the
setup for several different positions of source and first
mirror. Figure~\ref{fig:polcurve} shows how the intensity of the
detected light changes with the rotation angle. \citet{2013SPIE.8861E..1DM} perform this experiment in the laboratory albeit with different physical dimensions. In the figure, we compare the normalized simulated flux with the experimental results and find an excellent agreement. The statistical errors on the data are smaller than the plot symbols, but there is an additional systematic error in the measurement of the rotation angle (x-axis of the plot). The size of this error is not quantified in \citet{2013SPIE.8861E..1DM}, but large enough to explain the differences between simulated and measured data (Marshall, personal communication).

\subsection{Mission proposals}
\label{sect:use}
MARXS has already been used in \citet{2016SPIE.9905E..56G} to
demonstrate that a design based on critical-angle transmission
gratings \citep{2015SPIE.9603E..14H} can achieve a spectral
resolving power of several thousand with very modest requirements on
the grating alignment using the baseline design for Lynx
\citep{2015SPIE.9601E..0JG} and ARCUS
\citep{doi:10.1117/12.2062671}. ARCUS is currently proposed as an
explorer mission to NASA and Lynx is one of the studies for NASA's 2020
decadal survey. Particularly for ARCUS, simulations with MARXS were
instrumental to define the optical layout.

The capability for polarization ray-tracing allowed refinements of the
optical design for the REDSoX mission (Marshal et al., in prep), a
proposed sounding rocket experiment for a soft X-ray polarimeter.

\section{Summary}
\label{sect:summary}
MARXS is a new ray-tracing code in python that is designed to simulate the
performance of X-ray instrumentation on the ground and in space. It includes
modules for common elements such as mirrors, dispersion gratings, and
detectors. MARXS tracks the probability of a photon to travel along its
pathway without absorption by the optical elements it passes. Polarization is
taken into account where appropriate. We show the use of MARXS to study
sub-aperturing, to replicate the laboratory setup of an X-ray beamline, and to
design future instrumentation.

\acknowledgments 
We thank the Astropy collaboration for the set of tools and
helper packages that made the development of MARXS faster and easier, in
particular T.\ Robitaille and B.\ Sipocz for their automated updates of the package template. Support for this work was
provided by the National Aeronautics and Space Administration through the
Smithsonian Astrophysical Observatory contract SV3-73016 to MIT for Support of
the Chandra X-Ray Center, which is operated by the Smithsonian Astrophysical
Observatory for and on behalf of the National Aeronautics Space Administration
under contract NAS8-03060. JF and ATS also supported by NASA APRA
grant NNX17AE11G to work on parts of this code.

\software{astropy \citep{2013A&A...558A..33A},
 numpy \citep{numpy}
          }

\bibliographystyle{../AAStex/v611/aasjournal}
\bibliography{../articles}


\end{document}